# Vulnerable Source Code Detection using SonarCloud Code Analysis

Alifia Puspaningrum[1][a], Muhammad Anis Al Hilmi*[1][b], Darsih[1], Muhamad Mustamiin[1], Maulana Ilham Ginanjar[1]
[1]Department of Informatics, Politeknik Negeri Indramayu, Jalan Lohbener Lama No. 08, Indramayu, Indonesia
alifia.puspaningrum@polindra.ac.id, alhilmi@polindra.ac.id, darsih@poindra.ac.id, mustamiin@polindra.ac.id, ilhamginanjar24@gmail.com

Abstract: In Software Development Life Cycle (SDLC), security vulnerabilities are one of the points introduced during the construction stage. Failure to detect software defects earlier after releasing the product to the market causes higher repair costs for the company. So, it decreases the company's reputation, violates user privacy, and causes an unrepairable issue for the application. The introduction of vulnerability detection enables reducing the number of false alerts to focus the limited testing efforts on potentially vulnerable files. UMKM Masa Kini (UMI) is a *Point of Sales* application to sell any Micro, Small, and Medium Enterprises Product (UMKM). Therefore, in the current work, we analyze the suitability of these metrics to create Machine Learning based software vulnerability detectors for UMI applications. Code is generated using a commercial tool, SonarCloud. Experimental result shows that there are 3,285 vulnerable rules detected.

## 1 INTRODUCTION

In Software Development Life Cycle (SDLC), security vulnerabilities being one of point introduced during construction stage. However, vulnerabilities being one of issue which difficult to be detected until it becomes apparent as security failures in the operational stage of SDLC [1]. Failure to detect software defect earlier after releasing product to the market causes higher repairing cost for the company. So, it decreases company reputation, violate user privacy, and cause unrepairable issue for the application [2]. In addition, techniques to detect software vulnerabilities are needed before releasing product [3]. To solve those problems, dedicated tools are available on the market: e.g., Veracode [4] and SonarCode [5]. The introduction of vulnerability detection (usually a binary classification of vulnerable and neutral parts of the source code) enables reducing the number of false alerts to focus the limited testing efforts on potentially vulnerable files [6].

UMKM Masa Kini (UMI) is a *Point of Sales* application to sell any kind Micro, Small and Medium Enterprises Product (UMKM). Not only selling the products, UMI also provides facilities for offline transaction and facilities which can support the development of UMKM. However, in construction process, automated testing to detect vulnerable code is a good way to save money and time. Therefore, in the current work, we perform an analysis of the suitability of these metrics to create Machine Learning based software vulnerability detectors for UMI application. Code is generated using a commercial tool, SonarCloud.

## 2 LITERATURE REVIEW

### 2.1 Software Testing

Testing is an activity to evaluate software quality and to improve it [7]. In general, testing divided into two namely black box and white box testing. White box is a testing technique that uses Software Under Test (SUT) or the software being tested as a test guide to be carried out [8]. In addition, Black Box Testing is not an alternative solution to White Box Testing but is more of a complement to testing things that are not covered by White Box Testing.

---

[a]  https://orcid.org/0000-0001-7179-8847
[b]  https://orcid.org/0000-0003-3696-0807 (corresponding author)

## 2.2 Software Quality ISO 25010

Software quality can be assessed through certain metrics, methods, as well as through software tests. One of benchmarks for software quality is (International Organization for Standardization) ISO 25010. ISO/IEC 25010 has eight characteristics to measure software quality, including portability, performance efficiency, reliability, security usability, maintainability, compatibility, and functional suitability.

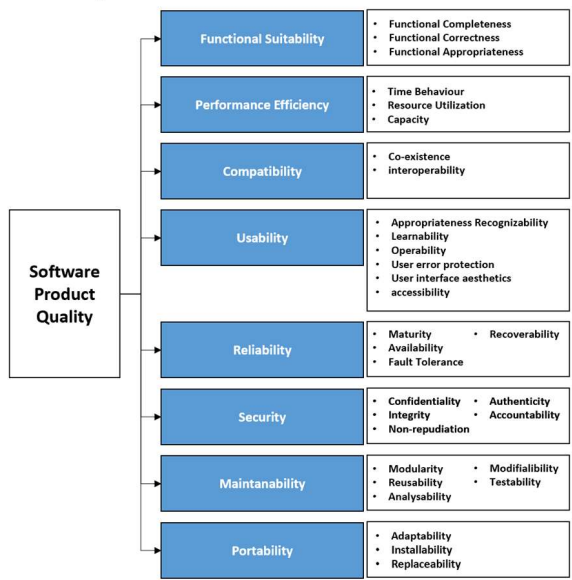

Figure 1. Product Quality Model ISO/IEC 25010

Figure 1 is characteristics and sub-characteristics of Software product quality model which consists of:

1. Functional Suitability is a characteristic to measure which system provides functions according to its requirement when used in certain conditions.
2. Performance Efficiency is a characteristic to calculate relative performance of the resources when used in certain conditions.
3. Compatibility is a characteristic to measure how a system share an information to other systems and execute required functions while sharing same hardware or software environment.
4. Usability is a characteristic to measure which system can be used to achieve the specified goals effectively and efficiently.
5. Reliability is a characteristic to measure how reliable a system can execute functions under specified conditions for a certain period.
6. Security is a characteristic to measure a system in protecting information and data, so that the system has a data access level according to the type and level of authorization.
7. Maintainability is a characteristic to represent the level of effectiveness and efficiency in the modification process for system improvement in accordance with adjustments and changes in the operational environment.
8. Portability is a characteristic to represent the level of effectiveness and efficiency of the system in transferring from one device to another.

## 2.3 SonarCloud

SonarCloud is an online automated software quality analysis platform delivered by SonarQube, which is used to collect code quality analysis data on linked software from GitHub [9] SonarCloud calculates several metrics such as number of lines of code and code complexity and verifies code compliance with a specific set of "coding rules" defined for most common development languages.

If the analyzed source code violates the coding rules or if the metric falls outside a predefined threshold, SonarCloud generates an "issue". SonarCloud includes 3 quality models from ISO 25010 namely Reliability, Maintainability and Security.

SonarCloud also classifies rules into five levels of code severity, namely Blocker, Critical, Major, Minor, and Info.

- Blocker: a bug with a high probability of affecting the use of the application in production.
- Critical: a bug with a low probability of affecting application use in production or a weak level of security.
- Major: lack of quality which can have a huge impact on developer productivity. Like code snippets not found and unused parameters.
- Minor: lack of quality which can slightly impact developer productivity. Like too long lines of code and minimal statements
- Info: not a bug or lack of quality but just a finding from SonarCloud.

# 3 METHOD

## 3.1 SonarCloud Pipeline

For experimental analysis, SonarCloud is used to collect vulnerable source code to obtain analysis of vulnerable code. Figure 2. show how SonarCloud detects insecure in given code.

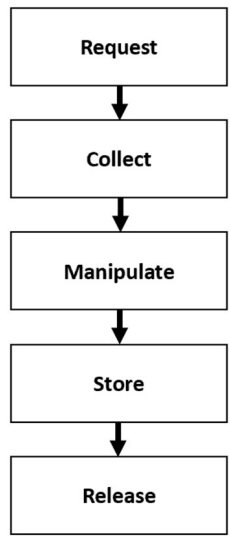

Figure 2. SonarCloud Block Diagram

The first step is request SonarCloud REST API. In the second phase, the crawlers collect source code of the application. In the next phase, SonarCloud manipulate by adding vulnerable lines as comments at the end of the source code file. After that, SonarCloud stores the result in the local file of the system. The last step is releasing the dataset to the community.

SonarCloud inspects the model according to ISO 25010 Standard, namely: *reliability, maintainability, security* and classify into five categories then.

## 3.2 Dataset

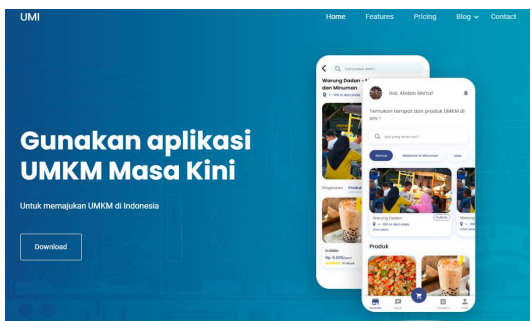

Figure 3. Landing Page of UMI Website

UMKM Masa Kini (UMI) is marketplace application which store and generate finance administration. This application has been developed since 2020. There are 19 functions in UMI.

# 4 RESULT

After scanning using SonarCloud as shown in Figure 4, SonarCloud generate a report, showing three scopes of the application namely *Reliability*, *Maintainability*, *Security* and its severity level as shown in Figure 5.

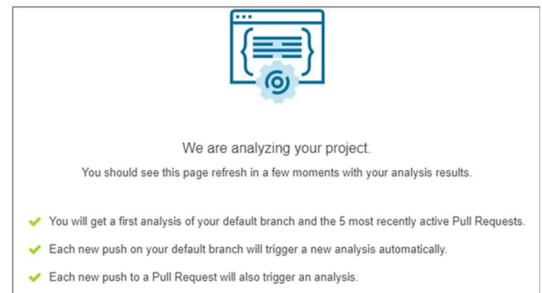

Figure 4. Analyzing Project Process

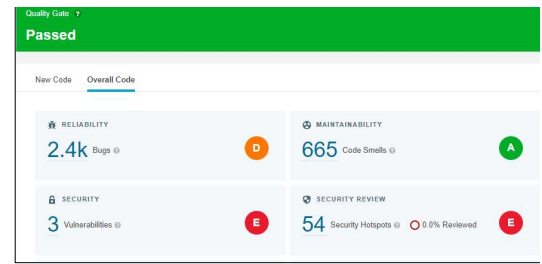

Figure 5. SonarCloud Report Analysis

Overall code rating (reliability, maintainability, security). The detail information of the report is shown in Table 1.

**Table 1** UMI Severity Level

| Type | Severity | Rules |
|---|---|---|
| Bug | All | 2,370 |
| | Info | - |
| | Minor | 2,000 |
| | Major | 364 |
| | Critical | 6 |
| | Blocker | - |
| Code Smell | All | 665 |
| | Info | - |
| | Minor | 74 |
| | Major | 356 |

| Type | Severity | Rules |
|---|---|---|
| | Critical | 233 |
| | Blocker | 2 |
| | All | 3 |
| | Info | - |
| Vulnerability | Minor | - |
| | Major | - |
| | Critical | 1 |
| | Blocker | 2 |
| Total | | 3,285 |

After retrieving the source code, for each vulnerability, a line is appended to the original source. Result shows that there are 3.285 rules detected as vulnerable code. Each issue namely bug, code smell, and vulnerability are categorized into severity level, namely: blocker, critical, major, minor, info. So that, developer can repair the code to get a safe code.

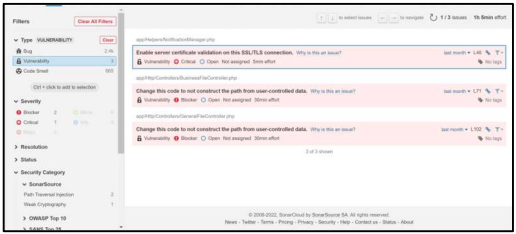

Figure 6. Vulnerability Analysis

There are three rules detected in vulnerability issue. As shown as Figure 6, three rules are reported and completed with each severity level. For each rule, SonarCloud describes the issue as shown as Figure 7.

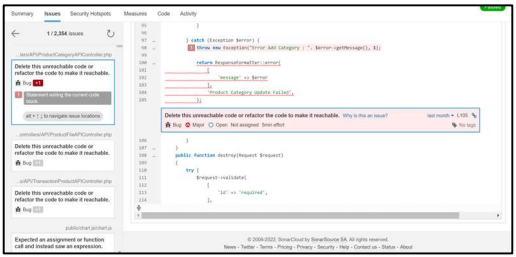

Figure 7 sample code of vulnerability type

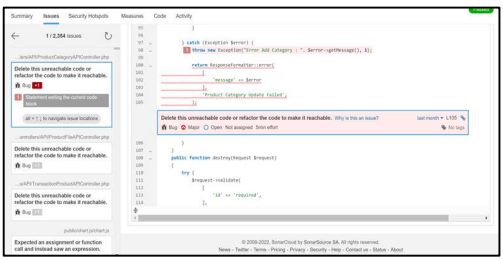

Figure 8 sample code of bug type

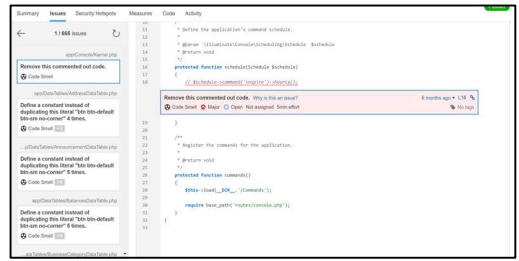

Figure 9 sample code of code smell type

For other issues such as code smell and bug category are also analyzed as shown as Figure 8 and Figure 9. Every report can help developer to do maintenance more efficient. However, one of limited of SonarCloud's web API is the one commonly known as the 10,000-issue limit. This constraint implies that every single request made will be responded to only with the first 10,000 results [10].

However, SonarCloud reports starting and ending line and starting and ending offset (column) of the vulnerability instead highlighting the vulnerable code.

## 5    CONCLUSIONS

This paper analyzes vulnerable for UMI application source code by using SonarCloud. Experimental result shows that there are 3,285 vulnerable rules detected. So that, developer can repair the code to get a safe code. For the future work, highlighting the vulnerable code instead of starting and ending line and starting and ending offset (column) of the vulnerability.

## ACKNOWLEDGEMENTS

This research has been supported by Pusat Penelitian dan Pengabdian Masyarakat, Politeknik Negeri Indramayu, as Penelitian Kerjasama Perguruan Tinggi 2022.